\documentclass[10pt,final,journal,letter,oneside,twocolumn]{IEEEtran}

\usepackage{cite}
\usepackage{ulem}
\usepackage{soul,color}
\usepackage{srcltx}
\usepackage{multirow}
\usepackage[tbtags,sumlimits,nointlimits,reqno]{amsmath}
\usepackage{graphicx,dblfloatfix}
\usepackage{float}
\usepackage{subfigure}
\usepackage{psfrag}
\usepackage{color}
\usepackage{amssymb}
\usepackage{makecell}
\usepackage{array}

\begin{document}

\title{Coupling Matrix Synthesis \\ of Nonreciprocal Lossless Two-Port Networks \\ using Gyrators and Inverters}

\author{Qingfeng~Zhang,~\IEEEmembership{Member,~IEEE,}
        Toshiro~Kodera,~\IEEEmembership{Senior Member,~IEEE,}
        Shuai Ding,~\IEEEmembership{Member,~IEEE,}
        and~Christophe~Caloz,~\IEEEmembership{Fellow,~IEEE}
\thanks{Manuscript received in 2014.}
\thanks{Q. Zhang is with the Department of Electronics and Electrical Engineering, South University of Science and Technology of China, Shenzhen, Guangdong, China 518055 (email: zhang.qf@sustc.edu.cn).}
\thanks{S. Ding and C. Caloz are with the Department of Electrical Engineering, Poly-Grames Research Center,
        \'{E}cole Polytechnique de Montr\'{e}al, Montr\'{e}al, QC, Canada H3T 1J4. C. Caloz is also with King Abdulaziz University, Jeddah, Saudi Arabia (email: christophe.caloz@polymtl.ca).}
\thanks{T. Kodera is with Meisei University, Hino, Tokyo 191-8506, Japan. (email: toshiro.kodera@meisei-u.ac.jp).}
}

\markboth{IEEE TRANSACTIONS ON MICROWAVE THEORY AND TECHNIQUES}%
{Shell \MakeLowercase{\textit{et al.}}: Bare Demo of IEEEtran.cls for Journals}
\maketitle


\begin{abstract}
A coupling matrix technique is presented for the synthesis of nonreciprocal lossless two-port networks. This technique introduces complex inverters to build the nonreciprocal transversal network corresponding to the nonreciprocal coupling matrix. It subsequently transforms this matrix into canonical topologies through complex similarity transformations. The complex inverters in the final topology are transformed into real inverters and gyrators for implementation simplicity. A second-order example is then given to illustrate the proposed technique. Possible implementations of the gyrators involved in the final design are also discussed.
\end{abstract}
\IEEEoverridecommandlockouts

\IEEEpeerreviewmaketitle

\section{Introduction}

The coupling matrix technique, originally proposed in the early 1970s by Atia and Williams~\cite{atia1971new,atia1972narrow} and later extensively extended and promoted by Cameron~\cite{cameron2007microwave}, is a powerful tool to synthesize microwave filters with quasi-elliptic responses.

The coupling matrix technique is in fact a general network technique, and its applications extend well beyond the area of pure filters. It was applied to the synthesis of multi-port networks, such as power dividers~\cite{cmx_multiport}, diplexers~\cite{cmx_multiport,cmx_diplexer} and MIMO antenna decoupling networks~\cite{antenna_decoupling_KLWU}. It was also applied to multi-function antennas~\cite{filtering_antenna} and cross-coupled phasers~\cite{Zhang_TMTT_cross} for microwave analog signal processing~\cite{caloz2013mm}.

It seems that all the schemes involving the coupling matrix technique reported to date have been limited to \emph{reciprocal networks}. In the area of nonreciprocal networks, the only synthesis technique reported consisted in directly cascading networks including gyrators~\cite{rubin1962cascade} to realize nonreciprocal lossless networks~\cite{tellegen1948gyrator}. In the present paper, we extend the coupling matrix synthesis technique to lossless \emph{nonreciprocal networks}. To the best knowledge of the authors, this represents the first report on nonreciprocal coupling matrix synthesis.

The main contributions of the paper are the following: 1)~Derivation of the skew symmetry condition for the admittance matrix of nonreciprocal (including the particular case of reciprocal) lossless networks and application of this condition to show that the unitary property condition of the scattering matrix still holds in the nonreciprocal case (Sec.~II); 2)~Demonstration that a lossless nonreciprocal network is magnitude-wise reciprocal and only phase-wise nonreciprocal (Sec.~III); 3)~Extension of the coupling matrix technique to the non-reciprocal case (Sec.~IV), including the introduction of nonreciprocal complex inverters (Sec.~IV-C), the derivation of the nonreciprocal coupling matrix involving these inverters (Sec.~IV-D), the development of complex similarity transformation for generalized topology transformation (Sec.~IV-E), and an additional similarity transformation to reduce complex couplings into real coupling (Sec.~IV-F). Moreover, a complete illustrative example is provided (Sec.~V), as well as a general discussion about the possible implementation of required gyrators, a comparison between real inverter and gyrators and an alternative implementation of complex inverters.

\section{General -- Possibly Nonreciprocal -- \\ Lossless Network}
\subsection{Admittance Matrix Condition}
The common condition for losslessness of a network, generally given in microwave textbooks (e.g.~\cite{pozar2009microwave}), is that the admittance matrix be purely imaginary. However, this condition is derived for reciprocal lossless networks and does not apply to nonreciprocal lossless networks. A more general lossless condition, valid for both reciprocal and nonreciprocal networks, is therefore derived next.

Consider an $N-$port network with port voltage phasor vector $[V]$ and current phasor vector $[I]$. The time average power delivered to the network is
  \begin{equation}\label{eq:power}
    P=\frac{1}{2}[V]^t[I]^*=\sum_{m=1}^NV_mI_{m}^*.
  \end{equation}
\noindent Substituting Ohm law, $[I]=[Y][V]$, where $[Y]$ is the admittance matrix of the network, into~\eqref{eq:power} yields
  \begin{equation}\label{eq:power2}
    P=\frac{1}{2}[V]^t[Y]^*[V]^*=\frac{1}{2}\sum_{n=1}^N\sum_{m=1}^NV_mY_{mn}^*V_n^*,
  \end{equation}
\noindent where $V_m$ and $V_n$ are the $m^\text{th}$ and $n^\text{th}$ elements of $[V]$, respectively, while $Y_{mn}$ is the element of the $m^\text{th}$ row and $n^\text{th}$ column of $[Y]$.

If the network is lossless then its net power (some of incident and refracted powers at all ports) is zero~\cite{pozar2009microwave}, i.e.
  \begin{equation}\label{eq:power_lossless}
    \text{Re}\{P\}=0.
  \end{equation}
Upon substitution of~\eqref{eq:power2}, the left-hand side of~\eqref{eq:power_lossless} may developed as
  \begin{align}\label{eq:real_power}
  \text{Re}\{P\}&=\frac{P+P^*}{2} \nonumber \\
  &=\frac{1}{4}\left(\sum_{n=1}^N\sum_{m=1}^NV_mY_{mn}^*V_n^*+\sum_{n=1}^N\sum_{m=1}^NV_m^*Y_{mn}V_n\right)\nonumber \\
  &=\frac{1}{4}\left(\sum_{n=1}^N\sum_{m=1}^NV_mY_{mn}^*V_n^*+\sum_{n=1}^N\sum_{m=1}^NV_mY_{nm}V_n^*\right)\nonumber \\
  &=\frac{1}{4}\sum_{n=1}^N\sum_{m=1}^NV_m\left(Y_{mn}^*+Y_{nm}\right)V_n^* \nonumber \\
  &=\frac{1}{2}[V]^t\left([Y]^*+[Y]^t\right)[V]^*.
  \end{align}
\noindent Since $[V]$ is external to the network and can therefore be chosen arbitrarily, $([Y]^*+[Y]^t)$ must be zero to satisfy~\eqref{eq:power_lossless}, i.e.
  \begin{equation}\label{eq:Y_condition}
    [Y]^*+[Y]^t=0,
  \end{equation}
\noindent which indicates that $[Y]$ is \emph{skew Hermitian symmetric} ($Y_{mn}=-Y_{nm}^*$).

The skew Hermitian symmetry condition~\eqref{eq:Y_condition} has been derived as a \emph{completely general condition for a lossless network}, and therefore also applies to nonreciprocal networks. Note that the conventional purely imaginary condition~\cite{pozar2009microwave} is only a particular case of~\eqref{eq:Y_condition}, corresponding to the reciprocal network with $[Y]^t=[Y]$.

This condition will be used for the construction of lossless complex inverters in Sec.~\ref{sec:complex_inverter}.

\subsection{Scattering Matrix Condition}

The scattering matrix $[S]$ is related to the admittance matrix $[Y]$ by~\cite{pozar2009microwave}
  \begin{equation}\label{eq:Y_S}
    [Y]=([U]-[S])([U]+[S])^{-1},
  \end{equation}
\noindent where $[U]$ is the identity matrix. Substituting~\eqref{eq:Y_S} into~\eqref{eq:Y_condition} yields
  \begin{equation}\label{eq:S_insert}
  ([U]+[S]^t)^{-1}([U]-[S]^t)+([U]-[S]^*)([U]+[S]^*)^{-1}=0.
  \end{equation}
\noindent Premultiplying by $([U]+[S]^t)$ and postmultiplying by $([U]+[S]^*)$ transforms \eqref{eq:S_insert} into
\begin{equation}\label{eq:S_multiply}
  ([U]-[S]^t)([U]+[S]^*)+([U]+[S]^t)([U]-[S]^*)=0.
  \end{equation}
\noindent Finally, upon expansion and simplification, \eqref{eq:S_multiply} becomes
  \begin{equation}\label{eq:S_condition}
  [S]^t[S]^*=[U].
  \end{equation}
\noindent Note that the unitary condition \eqref{eq:S_condition} has been derived as a \emph{completely general condition for a lossless network}, and therefore also applies to nonreciprocal networks, since~\eqref{eq:S_condition} is based on~\eqref{eq:Y_condition} which covers the nonreciprocal case, although it is most often derived in the particular case of reciprocal lossless networks~\cite{pozar2009microwave}.

\section{Nonreciprocal Lossless Two-Port Network}

In most cases, and particularly in the context of the coupling matrix technique, we are interested in two-port networks. Therefore, the lossless condition derived above for multi-port networks will be specialized here to two-port networks.

We shall prove that in nonreciprocal lossless two-port networks the magnitude of the transfer function is always reciprocal, so that only the phase of the transfer function is nonreciprocal.

In order to be lossless, the two-port network must satisfy~\eqref{eq:S_condition}, i.e.
  \begin{equation}\label{eq:s_unitary}
  {\left[ {\begin{array}{*{20}{c}}
  {{S_{11}}}&{{S_{12}}}\\
  {{S_{21}}}&{{S_{22}}}
  \end{array}} \right]^t}{\left[ {\begin{array}{*{20}{c}}
  {{S_{11}}}&{{S_{12}}}\\
  {{S_{21}}}&{{S_{22}}}
  \end{array}} \right]^*} = \left[ {\begin{array}{*{20}{c}}
  1&0\\
  0&1
  \end{array}} \right],
  \end{equation}
\noindent which may be expanded as
  \begin{subequations}\label{eq:s_expansion}
  \begin{align}
    &S_{11}S_{11}^*+S_{21}S_{21}^*=1,\label{eq:s_expansion_a}\\
    &S_{12}S_{12}^*+S_{22}S_{22}^*=1,\label{eq:s_expansion_b}\\
    &S_{11}S_{12}^*+S_{21}S_{22}^*=0,\label{eq:s_expansion_c}\\
    &S_{12}S_{11}^*+S_{22}S_{21}^*=0.\label{eq:s_expansion_d}
  \end{align}
  \end{subequations}
\noindent Note that the first three equations in~\eqref{eq:s_expansion} are independent from each other while the fourth can be derived from them (e.g. by taking the complex conjugate of the third one). So, this equation may be ignored in forthcoming derivations.

One may express $S_{11}$ in terms of the other scattering parameters by reformulating~\eqref{eq:s_expansion_c} as
  \begin{equation}\label{eq:s11}
  S_{11}=-\dfrac{S_{22}^*S_{21}}{S_{12}^*}.
  \end{equation}
\noindent Inserting~\eqref{eq:s11} into~\eqref{eq:s_expansion_a} leads to
  \begin{equation}\label{eq:s11_inserting}
  \dfrac{S_{22}S_{22}^*S_{21}S_{21}^*}{S_{12}S_{12}^*}+S_{21}S_{21}^*=1.
  \end{equation}
Multiplying both sides of~\eqref{eq:s11_inserting} by $S_{12}S_{12}^*$ and rearranging the result, one obtains
  \begin{equation}\label{eq:s11_reform}
  (S_{22}S_{22}^*+S_{12}S_{12}^*)S_{21}S_{21}^*=S_{12}S_{12}^*.
  \end{equation}
\noindent Inserting~\eqref{eq:s_expansion_b} into~\eqref{eq:s11_reform} finally leads to
  \begin{equation}\label{eq:s21_recip}
  S_{21}S_{21}^*=S_{12}S_{12}^*,
  \end{equation}
\noindent or
  \begin{equation}\label{eq:s21_mag}
  |S_{21}|=|S_{12}|,
  \end{equation}
\noindent which reveals that \emph{the magnitude of the transfer function of a lossless two-port network is necessarily reciprocal}. However, no constraint is found on the phase of the transfer function, which has therefore to be the nonreciprocal part of the network. So, in general (except in the reciprocal case), we have
  \begin{equation}\label{eq:s21_angle}
  \angle S_{21}\neq\angle S_{12}.
  \end{equation}

The relation between the magnitudes of $S_{11}$ and $S_{22}$ is obtained by subtracting \eqref{eq:s_expansion_b} from~\eqref{eq:s_expansion_a} and using~\eqref{eq:s21_mag}, which results into
  \begin{equation}\label{eq:s22_s11}
  |S_{11}|=|S_{22}|,
  \end{equation}
\noindent while no restriction is found on the phases of $S_{11}$ and $S_{22}$, so that in general
  \begin{equation}\label{eq:phase_s22_s11}
  \angle S_{11}\neq \angle S_{22}.
  \end{equation}

Equations~\eqref{eq:s21_mag} to \eqref{eq:phase_s22_s11} have derived for the particular case of two-port networks. However, these results generalize to $N>2$ multi-port networks, i.e. $|S_{ij}|=|S_{ji}|$ $\forall{i,j}=1,2,\ldots{N}$, and generally \mbox{$\angle{S_{ij}}\neq\angle{S_{ji}}$} and $\angle{S_{ii}}\neq\angle{S_{jj}}$ $\forall{i,j}=1,2,\ldots{N}$ $i\neq j$ .

\section{Coupling Matrix Synthesis for a Nonreciprocal Lossless Two-Port Network}
\subsection{Minimum-Order Coupling Matrix}

The order of a coupling matrix is directly proportional to the size of the corresponding device implementation since it corresponds to the number of resonators. Therefore, one generally seeks for minimum-order coupling matrices. Since the order of the coupling matrix is the same as the order of the denominator polynomial of the rational function forming the admittance parameter, minimizing the coupling matrix order reduces to minimizing the denominator order of the admittance function. This section will derive the corresponding conditions.

The admittance parameters $Y_{21}$ and $Y_{22}$ are related to the scattering parameters by the formulas~\cite{pozar2009microwave}
  \begin{subequations}\label{eq:y_conversion}
  \begin{align}
    &Y_{21}=\dfrac{-2S_{21}}{(1+S_{11})(1+S_{22})-S_{12}S_{21}},\label{eq:y_conversion_a}\\
    &Y_{22}=\dfrac{(1+S_{11})(1-S_{22})+S_{12}S_{21}}{(1+S_{11})(1+S_{22})-S_{12}S_{21}},\label{eq:y_conversion_b}
  \end{align}
  \end{subequations}
\noindent whose denominator can be simplified by using~\eqref{eq:s11} and~\eqref{eq:s_expansion_b} as
  \begin{align}\label{eq:denominator}
  &(1+S_{11})(1+S_{22})-S_{12}S_{21}\nonumber\\
  &=1+S_{11}+S_{22}+S_{11}S_{22}-S_{12}S_{21}\nonumber\\
  &=1+S_{11}+S_{22}-\dfrac{S_{22}^*S_{21}S_{22}}{S_{12}^*}-S_{12}S_{21}\nonumber\\
  &=1+S_{11}+S_{22}-\dfrac{S_{21}}{S_{12}^*}(S_{22}S_{22}^*+S_{12}S_{12}^*)\nonumber\\
  &=1+S_{11}+S_{22}-\dfrac{S_{21}}{S_{12}^*}.
  \end{align}
\noindent The numerator of~\eqref{eq:y_conversion_b} can be similarly simplified to

 \begin{align}
  &(1+S_{11})(1-S_{22})+S_{12}S_{21}
  =1+S_{11}-S_{22}+\dfrac{S_{21}}{S_{12}^*}.
  \end{align}
With these simplifications, Eqs.~\eqref{eq:y_conversion} become
  \begin{subequations}\label{eq:y_conversion2}
  \begin{align}
    &Y_{21}=\dfrac{-2S_{21}}{1+S_{11}+S_{22}-\dfrac{S_{21}}{S_{12}^*}},\label{eq:y_conversion2_a}\\
    &Y_{22}=\dfrac{1+S_{11}-S_{22}+\dfrac{S_{21}}{S_{12}^*}}{1+S_{11}+S_{22}-\dfrac{S_{21}}{S_{12}^*}}.\label{eq:y_conversion2_b}
  \end{align}
  \end{subequations}

The scattering matrix of a two-port network is usually expressed in terms of ratios of polynomials as
  \begin{equation}\label{eq:s_polynomial}
  {\left[ {\begin{array}{*{20}{c}}
  {{S_{11}}}&{{S_{12}}}\\
  {{S_{21}}}&{{S_{22}}}
  \end{array}} \right]} = \frac{1}{H(s)}\left[ {\begin{array}{*{20}{c}}
  {{F_{11}(s)}}&{{P_{12}(s)}}\\
  {{P_{21}(s)}}&{{F_{22}(s)}}
  \end{array}} \right],
  \end{equation}
\noindent where $s=j\omega+\sigma$ is the complex frequency and where $H(s)$ is a Hurwitz polynomial~\cite{cameron2007microwave}. Upon substitution of~\eqref{eq:s_polynomial}, Eq.~\eqref{eq:y_conversion2} becomes
  \begin{subequations}\label{eq:y_conversion3}
  \begin{align}
    &Y_{21}(s)=\dfrac{-2P_{21}(s)}{H(s)+F_{11}(s)+F_{22}(s)-\dfrac{P_{21}(s)}{P_{12}(s)^\star}H(s)^\star},\label{eq:y_conversion3_a}\\
    &Y_{22}(s)=\dfrac{H(s)+F_{11}(s)-F_{22}(s)+\dfrac{P_{21}(s)}{P_{12}(s)^\star}H(s)^\star}{H(s)+F_{11}(s)+F_{22}(s)-\dfrac{P_{21}(s)}{P_{12}(s)^\star}H(s)^\star}.\label{eq:y_conversion3_b}
  \end{align}
  \end{subequations}
where the symbol `$^\star$' represents a modified complex conjugate operation defined as $f(s)^\star=f^\text{`$*$'}(-s)$ with `$*$' meaning complex conjugate applying to the coefficients of $s$ powers but not to $s$ itself. This modified operator is necessary to ensure consistency between the physical quantities, namely the scattering parameters, which are complex functions of the real frequency $\omega$, and the mathematical quantities, namely the $P_{ij},F_{ii},H$ functions, which are complex functions of the complex variable~$s$.

Minimum order in the denominator of~\eqref{eq:y_conversion3} is obtained by setting $P_{21}(s)/P_{12}(s)^\star$ to be a constant independent of $s$ (zeroth-order rational function), i.e.
  \begin{equation}\label{eq:constant}
  \dfrac{P_{21}(s)}{P_{12}(s)^\star}=c.
  \end{equation}
\noindent Since $S_{21}(j\omega)$ and $S_{12}(j\omega)$ have equal magnitude, according to \eqref{eq:s21_mag}, the magnitude of their numerators must be equal as well for $s=j\omega$, i.e. $|P_{21}(j\omega)|$ and $|P_{12}(j\omega)|$, since they share the same denominator, $H(s)$\footnote{The order of the denominator may be larger if the right-hand side of~\eqref{eq:y_conversion3} depends on $s$, without violating $|P_{21}(j\omega)|=|P_{12}(j\omega)|$. For example, if $P_{21}(s)/P_{12}(s)^\star=(s-1)/(s+1)$, the denominator of \eqref{eq:y_conversion3} first becomes a rational function, and is transformed into a polynomial by multiplying it as well as the numerator [in~\eqref{eq:y_conversion3}] by $s+1$, which results in a denominator's order increased by one.}. Therefore, according to~\eqref{eq:constant}, the magnitude of the constant $c$ must be unity. In the most simple case, one may choose $c=\pm 1$. Then~\eqref{eq:constant} may be written as
  \begin{equation}\label{eq:mini_condition}
  P_{21}(s)=\pm P_{12}(s)^\star.
  \end{equation}
By inserting~\eqref{eq:mini_condition} and~\eqref{eq:s_polynomial} into~\eqref{eq:s11}, one also obtains
  \begin{equation}\label{eq:f11_f22}
  F_{11}(s)=\mp F_{22}(s)^\star.
  \end{equation}
Finally substituting~\eqref{eq:mini_condition} and~\eqref{eq:f11_f22} into~\eqref{eq:y_conversion3} yields
  \begin{subequations}\label{eq:y_conversion_final}
  \begin{align}
    &Y_{21}(s)=\dfrac{-2P_{21}(s)}{H(s)+F_{11}(s)\mp[H(s)+F_{11}(s)]^\star},\label{eq:y_conversion_final_a}\\
    &Y_{22}(s)=\dfrac{H(s)+F_{11}(s)\pm[H(s)+F_{11}(s)]^\star}{H(s)+F_{11}(s)\mp[H(s)+F_{11}(s)]^\star}.\label{eq:conversion_final_b}
  \end{align}
  \end{subequations}
The positive or negative signs in \eqref{eq:mini_condition}-\eqref{eq:y_conversion_final} should be chosen to make the denominator order of \eqref{eq:conversion_final_b} to be larger than the numerator order. One usually chooses $+$ sign in \eqref{eq:f11_f22} and the denominator of~\eqref{eq:y_conversion_final}, and $-$ sign in \eqref{eq:mini_condition} and the numerator of \eqref{eq:conversion_final_b}, when the order of $H(s)$ is even. The signs are reversed in the odd-order cases.

\subsection{Transversal Network}

In order to synthesize the network corresponding to the admittance polynomial functions~\eqref{eq:y_conversion_final}, one may factor out these functions\footnote{The reader might refer to Sec. V for a specific example of this operation.}, which leads to expressions of the form
  \begin{subequations}\label{eq:y_expansion}
  \begin{align}
    &Y_{21}(s)=r_{0}+\sum_{k=1}^N\dfrac{r_{21k}}{s-j\lambda_k},\label{eq:y_expansion_a}\\
    &Y_{22}(s)=\sum_{k=1}^N\dfrac{r_{22k}}{s-j\lambda_k},\label{eq:y_expansion_b}
  \end{align}
  \end{subequations}
\noindent where $N$ and $j\lambda_k$ are the order and $k^\text{th}$ root of the denominator of~\eqref{eq:y_conversion_final}, respectively, $r_{21k}$ and $r_{22k}$ are the $k$-th residues of $Y_{21}(s)$ and $Y_{22}(s)$, respectively, and $r_0$ is a complex constant which is nonzero only when the numerator and denominator of $Y_{21}(s)$ in~\eqref{eq:y_conversion_final_a} have the same order.

\begin{figure}[!t]
  \center
  \psfrag{1}[c][c]{\footnotesize $k=1$}
  \psfrag{2}[c][c]{\footnotesize $k=2$}
  \psfrag{3}[c][c]{\footnotesize $k=N-1$}
  \psfrag{4}[c][c]{\footnotesize $k=N$}
  \psfrag{5}[c][c]{\footnotesize \raisebox{.5pt}{\textcircled{\raisebox{-.9pt} {1}}}}
  \psfrag{6}[c][c]{\footnotesize \raisebox{.5pt}{\textcircled{\raisebox{-.9pt} {2}}}}
  \psfrag{S}[c][c]{\footnotesize S}
  \psfrag{L}[c][c]{\footnotesize L}

  \psfrag{a}[c][c]{\footnotesize \shortstack{Direct\\Coupling}}
  \includegraphics[width=6cm]{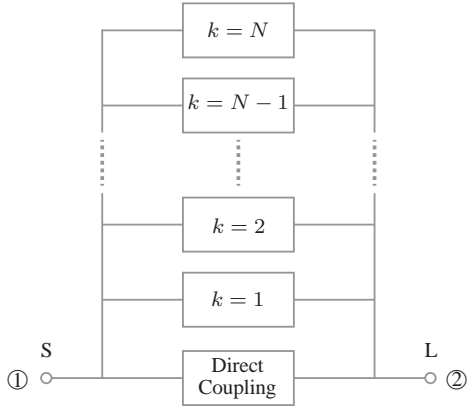}\\
  \caption{Transversal network corresponding to~\eqref{eq:y_expansion}. S and L represent the source and the load, respectively.}\label{fig:transversal}
\end{figure}

\begin{figure}[!t]
  \center
  \psfrag{1}[c][c]{\footnotesize $C_k$}
  \psfrag{2}[c][c]{\footnotesize $B_k$}
  \psfrag{3}[c][c]{\footnotesize $M_{Sk}$}
  \psfrag{4}[c][c]{\footnotesize $M_{kL}$}
  \includegraphics[width=7cm]{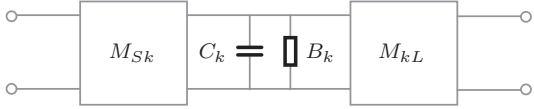}\\
  \caption{Conventional transversal element based on real admittance inverters ($M_{Sk}$ and $M_{kL}$) for a reciprocal network.}\label{fig:element_recip}
\end{figure}

The expanded admittance parameters in~\eqref{eq:y_expansion} correspond to the transversal network shown in Fig.~\ref{fig:transversal}, where the $k^\text{th}$ transversal element corresponds to the $k^\text{th}$ term of the sum~\eqref{eq:y_expansion} and the direct coupling between the source and load corresponds to $r_0$. It will be shown in the example of Sec.~\ref{sec_example} that $r_{21k}$ is complex under nonreciprocity, which is different from the reciprocal case where both $r_{22k}$ and $r_{21k}$ are real. The complex nature of $r_{21k}$ is thus a direct consequence of nonreciprocity and leads to a new transversal element configurations in Fig.~\ref{fig:transversal}. The conventional transversal element~\cite{cameron2007microwave}, shown in Fig.~\ref{fig:element_recip}, will then need to be replaced by a nonreciprocal complex inverter, as shall be shown next.

\subsection{Generalized Lossless Complex Inverter}
\label{sec:complex_inverter}

The generalized admittance matrix for a lossless complex inverter may be written
  \begin{equation}\label{eq:inverter_admittance}
  [Y]=j\left[ {\begin{array}{*{20}{c}}
  {{0}}&{{\widetilde{M}}}\\
  {{\widetilde{M}^*}}&{{0}}
  \end{array}} \right],
  \end{equation}
\noindent where $\widetilde{M}$ denotes a complex coupling coefficient with the tilde emphasizing the distinction with the conventional real coupling coefficient $M$ in Fig.~\ref{fig:element_recip}, and where a $j$ factor has been introduced for compliance with the conventional coupling matrix~\cite{cameron2007microwave}.

According to the definition of the admittance matrix, the voltages $[V_1,V_2]$ and currents $[I_1, I_2]$ should satisfy the relations
  \begin{subequations}\label{eq:V_I}
  \begin{align}
    &I_1=j\widetilde{M}V_2,\label{eq:V_I_a}\\
    &I_2=j\widetilde{M}^*V_1,\label{eq:V_I_b}
  \end{align}
  \end{subequations}
\noindent By multiplying the two equations~\eqref{eq:V_I} together and dividing the result by $V_1I_2$, one obtains
  \begin{align}
    I_1/V_1=\dfrac{|\widetilde{M}|^2}{-I_2/V_2},\label{eq:inversion_factor}
  \end{align}
 \noindent confirming that the admittance matrix in~\eqref{eq:inverter_admittance} operates as an admittance inverter with the inversion factor $|\widetilde{M}|^2$.

Note that the admittance matrix~\eqref{eq:inverter_admittance} satisfies the losslessness condition~\eqref{eq:Y_condition} which confirms that the inverter~\eqref{eq:inverter_admittance} is a lossless component. Also note that this inverter reduces to the conventional inverter when $\widetilde{M}$ is real and a gyrator~\cite{tellegen1948gyrator} when $\widetilde{M}$ is purely imaginary. As soon as $\widetilde{M}$ ceases to be purely real, the inverter becomes nonreciprocal.

The ABCD matrix of the generalized inverter \eqref{eq:inverter_admittance}, which will be needed next, is readily found as
  \begin{equation}\label{eq:inverter_abcd}
  \left[ {\begin{array}{*{20}{c}}
  {{0}}&{{\dfrac{j}{\widetilde{M}^*}}}\\
  {{j\widetilde{M}}}&{{0}}
  \end{array}} \right].
  \end{equation}
\subsection{Complex Coupling Matrix}

\begin{figure}[!t]
  \center
    \psfrag{1}[c][c]{\footnotesize $C_k$}
  \psfrag{2}[c][c]{\footnotesize $B_k$}
  \psfrag{3}[c][c]{\footnotesize $\widetilde{M}_{Sk}$}
  \psfrag{4}[c][c]{\footnotesize $M_{kL}$}
  \includegraphics[width=7cm]{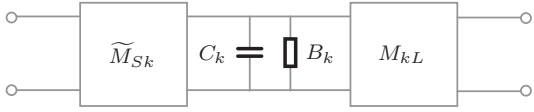}\\
  \caption{New transversal element including complex and real inverters for nonreciprocal networks.}\label{fig:element_nonrecip}
\end{figure}

Building the transversal network corresponding to the admittance parameters~\eqref{eq:y_expansion} may be done by introducing the transversal element in Fig.~\ref{fig:element_nonrecip}, where one of the two conventional inverters has been replaced by a complex inverter with ABCD matrix given by~\eqref{eq:inverter_abcd}. Note that one conventional inverter is still used. It will be shown later that the conventional inverter is maintained is because $r_{22k}$ in~\eqref{eq:y_expansion} is real.

\begin{figure}[!t]
  \center
  \psfrag{a}[c][c]{\footnotesize $\widetilde{M}_{S1}$}
  \psfrag{s}[c][c]{\footnotesize $\widetilde{M}^*_{S1}$}
   \psfrag{b}[c][c]{\footnotesize $\widetilde{M}_{S2}$}
  \psfrag{r}[c][c]{\footnotesize $\widetilde{M}^*_{S2}$}
  \psfrag{d}[c][c]{\scriptsize $\widetilde{M}_{SN-1}$}
  \psfrag{q}[c][c]{\scriptsize $\widetilde{M}^*_{SN-1}$}
  \psfrag{e}[c][c]{\footnotesize $\widetilde{M}_{SN}$}
  \psfrag{p}[c][c]{\footnotesize $\widetilde{M}^*_{SN}$}
  \psfrag{f}[c][c]{\footnotesize $\widetilde{M}_{SL}$}
  \psfrag{m}[c][c]{\footnotesize $\widetilde{M}^*_{SL}$}
  \psfrag{g}[c][c]{\footnotesize $M_{L1}$}
  \psfrag{h}[c][c]{\footnotesize $M_{L2}$}
  \psfrag{j}[c][c]{\scriptsize $M_{LN-1}$}
  \psfrag{k}[c][c]{\footnotesize $M_{LN}$}
  \psfrag{c}[c][c]{\footnotesize $jB_1$}
  \psfrag{i}[c][c]{\footnotesize $jB_2$}
  \psfrag{n}[c][c]{\footnotesize $jB_{N-1}$}
  \psfrag{o}[c][c]{\footnotesize $jB_N$}
  \includegraphics[width=8cm]{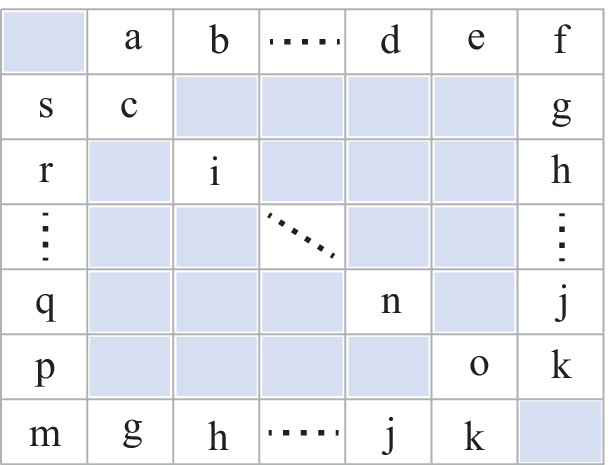}\\
  \caption{Complex coupling matrix, $[\widetilde{M}]$, corresponding to the transversal network in Fig.~\ref{fig:transversal} and element in Fig.~\ref{fig:element_nonrecip}. The shaded entries represent zeros.}\label{fig:cmx_transversal}
\end{figure}

The transversal network in Fig.~\ref{fig:transversal}, which is then composed of $N$ elements with the configuration in Fig.~\ref{fig:element_nonrecip}, corresponds to the $N+2$ coupling matrix shown in Fig.~\ref{fig:cmx_transversal}. Note that the coupling coefficients in the first row and first column are conjugate pairs of each other, corresponding to the input nonreciprocal complex inverters [Eq.~\eqref{eq:inverter_admittance}] of the transversal element in Fig.~\ref{fig:element_nonrecip}, whereas those in the last row and last column are equal quantities corresponding to the output conventional reciprocal inverters of the transversal element in Fig.~\ref{fig:element_nonrecip}. $\widetilde{M}_{SL}$ corresponds to direct complex coupling between the source (S) and the load (L).

To compute the entries of the matrix in Fig.~\ref{fig:cmx_transversal}, we shall calculate the corresponding circuital admittance parameters and compare them with their mathematical counterparts~\eqref{eq:y_expansion}. The ABCD matrix of the $k^\text{th}$ transversal element in Fig.~\ref{fig:element_nonrecip} is obtained as
  \begin{align}\label{eq:element_abcd}
  &\left[ {\begin{array}{*{20}{c}}
  {{0}}&{{\dfrac{j}{\widetilde{M}_{Sk}^*}}}\\
  {{j\widetilde{M}_{Sk}}}&{{0}}
  \end{array}} \right]  \left[ {\begin{array}{*{20}{c}}
  1&0\\
  C_ks+jB_k&1
  \end{array}} \right]  \left[ {\begin{array}{*{20}{c}}
  {{0}}&{{\dfrac{j}{M_{kL}}}}\\
  {{jM_{kL}}}&{{0}}
  \end{array}} \right]\nonumber\\
  &=\left[ {\begin{array}{*{20}{c}}
  {{-\dfrac{M_{kL}}{\widetilde{M}_{Sk}^*}}}&{{-\dfrac{C_ks+jB_k}{\widetilde{M}_{Sk}^*M_{kL}}}}\\
  {{0}}&{{-\dfrac{\widetilde{M}_{Sk}}{M_{kL}}}}
  \end{array}} \right],
  \end{align}
\noindent and the corresponding admittance parameters are obtained from this matrix by standard network transformation formulas~\cite{pozar2009microwave} as
  \begin{subequations}\label{eq:element_y}
  \begin{align}
    &Y_{21k}(s)=\dfrac{\widetilde{M}^*_{Sk}M_{kL}}{C_ks+jB_k},\label{eq:element_y_a}\\
    &Y_{22k}(s)=\dfrac{M^2_{kL}}{C_ks+jB_k}.\label{eq:element_y_b}
  \end{align}
  \end{subequations}
\noindent Note that the numerator of \eqref{eq:element_y_b} is purely real, consistently with the fact that the output inverter in Fig.~\ref{fig:element_nonrecip} is a conventional one and corresponding to real $r_{22k}$ in~\eqref{eq:y_expansion}. The admittance parameters of the overall transversal network is the sum of all its elements, i.e.
  \begin{subequations}\label{eq:whole_y}
  \begin{align}
    &Y_{21}(s)=j\widetilde{M}_{SL}+\sum_{k=1}^N\dfrac{\widetilde{M}^*_{Sk}M_{kL}}{C_ks+jB_k},\label{eq:whole_y_a}\\
    &Y_{22}(s)=\sum_{k=1}^N\dfrac{M^2_{kL}}{C_ks+jB_k}.\label{eq:whole_y_b}
  \end{align}
  \end{subequations}

Equating the network expressions~\eqref{eq:whole_y} and the mathematical expressions~\eqref{eq:y_expansion}, one obtains
  \begin{subequations}\label{eq:correspond}
  \begin{align}
    &C_k=1,\\
    &B_k=-\lambda_k,\\
    &\widetilde{M}_{SL}=-jr_0,\\
    &M_{kL}=\sqrt{r_{22k}},\\
    &\widetilde{M}_{Sk}=\dfrac{r_{21k}^*}{\sqrt{r_{22k}}}.
  \end{align}
  \end{subequations}

Equations~\eqref{eq:correspond} provide all the entries of the coupling matrix in Fig.~\ref{fig:cmx_transversal} corresponding to the transversal network of Fig.~\ref{fig:transversal}. However, this transversal network is only a temporary topology. It will be next transformed into more convenient topologies using similarity transformations.

\subsection{Topology Transformation}

\begin{figure}[!t]
  \center
  \psfrag{0}[c][c]{\footnotesize $1$}
  \psfrag{1}[c][c]{\footnotesize $1$}
  \psfrag{2}[c][c]{\footnotesize $i$}
  \psfrag{3}[c][c]{\footnotesize $j$}
  \psfrag{4}[c][c]{\footnotesize $N+2$}
  \psfrag{i}[c][c]{\footnotesize $c_r$}
  \psfrag{n}[c][c]{\footnotesize $c^*_r$}
  \psfrag{p}[c][c]{\footnotesize $s_r$}
  \psfrag{o}[c][c]{\footnotesize $-s^*_r$}
  \includegraphics[width=6cm]{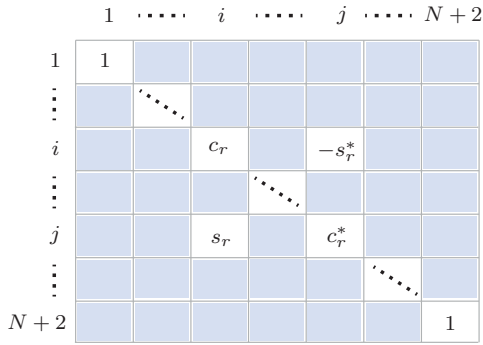}\\
  \caption{Complex rotation matrix $[R]$ for topology transformation ($s_r$ and $c_r$ are complex numbers, $2\leq i<j\leq N+1$).}\label{fig:transformation_topo}
\end{figure}

Topology transformation is accomplished via rotation operations of the coupling matrix~\cite{cameron2007microwave}. The rotations in~\cite{cameron2007microwave} are restricted to reciprocal networks and do not apply to nonreciprocal networks. Therefore, we introduce the new rotation operation
  \begin{align}
    [\widetilde{M}]^\prime=[R]\cdot [\widetilde{M}]\cdot ([R]^*)^t,\label{eq:rotation}
  \end{align}
\noindent where $[\widetilde{M}]$ and $[\widetilde{M}]^\prime$ are the original and rotated $(N+2)\times(N+2)$ coupling matrices, respectively, and $[R]$ is the rotation matrix shown in Fig.~\ref{fig:transformation_topo}. All the diagonal entries of $[R]$ except $[i,i]$ and $[j,j]$ are unity, and all the other (non-diagonal) entries are zero except $[i,j]$ and $[j,i]$. $[R]$ must be unitary to maintain~\eqref{eq:Y_condition}, and therefore the coefficients $s_r$ and $c_r$ must satisfy the condition
  \begin{align}
    c_rc_r^*+s_rs_r^*=1.\label{eq:cr_sr}
  \end{align}

Application of~\eqref{eq:rotation} changes only the entries in the $i^\text{th}$ and $j^\text{th}$ rows and columns of the matrix in Fig.~\ref{fig:cmx_transversal}, which transform as
\begin{subequations}\label{eq:after_rotation}
  \begin{align}
  \widetilde{M}_{ik}^\prime=c_r\widetilde{M}_{ik}-s_r^*\widetilde{M}_{jk},\\
  \widetilde{M}_{jk}^\prime=s_r\widetilde{M}_{ik}+c_r^*\widetilde{M}_{jk},\\
  \widetilde{M}_{ki}^\prime=c_r^*\widetilde{M}_{ki}-s_r\widetilde{M}_{kj},\\
  \widetilde{M}_{kj}^\prime=s_r^*\widetilde{M}_{ki}+c_r\widetilde{M}_{kj},
  \label{eq:after_rotation_b}
  \end{align}
 \end{subequations}
 \noindent where $1\leq k\leq N+2, k\neq i, k\neq j$.

In order to eliminate the entry $[k,j]$, one sets $\widetilde{M}_{kj}^\prime=0$ in~\eqref{eq:after_rotation_b}, which implies
  \begin{align}
\dfrac{s_r^*}{c_r}=-\dfrac{\widetilde{M}_{ki}}{\widetilde{M}_{kj}}=\sigma,\label{eq:ration_sr_cr}
  \end{align}
\noindent where $\sigma$ denotes $-{\widetilde{M}_{ki}}/{\widetilde{M}_{kj}}$. Inserting~\eqref{eq:ration_sr_cr} into~\eqref{eq:cr_sr} yields
  \begin{align}
c_rc_r^*=\dfrac{1}{1+\sigma\sigma^*},\label{eq:cr_cr}
  \end{align}
\noindent where no constraint is imposed on the phase of $c_r$.  For convenience, one may choose $c_r$ real, i.e.
  \begin{align}
c_r=\dfrac{1}{\sqrt{1+\sigma\sigma^*}}.\label{eq:cr}
  \end{align}
\noindent Inserting~\eqref{eq:cr} into~\eqref{eq:ration_sr_cr} leads to
  \begin{align}
s_r=\dfrac{\sigma^*}{\sqrt{1+\sigma\sigma^*}}.\label{eq:sr}
  \end{align}

The procedure to apply the topology transformation, is as follows: 1)~determines which $[k,j]$ entry is to be eliminated, 2)~calculate the corresponding value of $\sigma$ using~\eqref{eq:ration_sr_cr}, 3)~compute $c_r$ and $s_r$ using~\eqref{eq:cr} and~\eqref{eq:sr}, respectively, which provides the rotation matrix of Fig.~\ref{fig:transformation_topo}, 4)~apply~\eqref{eq:rotation} to generate the transformed coupling matrix $[M^\prime]$.

\subsection{Complex to Real Coupling Element Transformation}
\label{sec:comp_trans}

\begin{figure}[!t]
  \center
  \psfrag{0}[c][c]{\footnotesize $1$}
  \psfrag{1}[c][c]{\footnotesize $1$}
  \psfrag{2}[c][c]{\footnotesize $2$}
  \psfrag{3}[c][c]{\footnotesize $i$}
  \psfrag{4}[c][c]{\footnotesize $N+1$}
  \psfrag{5}[c][c]{\footnotesize $N+2$}
  \psfrag{i}[c][c]{\footnotesize $e^{j\theta}$}
  \includegraphics[width=6cm]{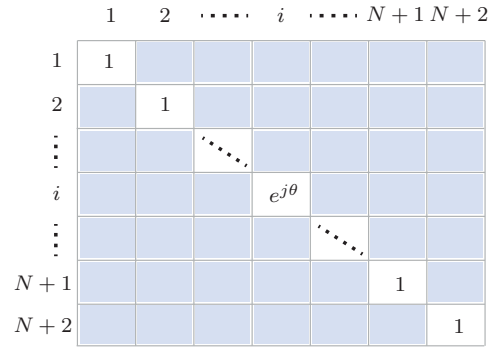}\\
  \caption{Complex transformation matrix $[R]$ for transforming complex coupling elements into real coupling elements.}\label{fig:transformation_complex}
\end{figure}

The rotation matrix in Fig.~\ref{fig:transformation_topo} can transform the transversal network in Fig.~\ref{fig:transversal} into a canonical network, such as for instance a folded or star network topology~\cite{cameron2007microwave}. However, most of the resulting coupling coefficients in the transformed network might generally still be complex in the case of nonreciprocity. Since such complex couplings imply higher complexity, they should be reduced to the smallest possible number. This may be accomplished using complex transformation matrix in Fig.~\ref{fig:transformation_complex}.

Applying~\eqref{eq:rotation} with the rotation matrix defined in Fig.~\ref{fig:transformation_complex} changes the $i$-th row and $i$-th column of $\left[\widetilde{M}\right]$ as
\begin{subequations}\label{eq:complex_transformed}
  \begin{align}
  &\widetilde{M}_{ik}^\prime=\widetilde{M}_{ik}e^{j\theta},\\
  &\widetilde{M}_{ki}^\prime=\widetilde{M}_{ki}e^{-j\theta},
  \end{align}
 \end{subequations}
\noindent where $1\leq k\leq N+2, k\neq i$, and $\widetilde{M}_{ik}$ and $\widetilde{M}_{ki}$ are conjugate pairs according to~\eqref{eq:inverter_admittance}, while leaving the rest of the matrix unaffected. Note that $\widetilde{M}_{ik}^\prime$ and $\widetilde{M}_{ki}^\prime$ are still conjugate pairs after the transformation.

To transform the complex coupling element between the $k$-th and $i$-th resonators, $\widetilde{M}_{ki}$, one first expresses it in polar form,
  \begin{align}
  \widetilde{M}_{ki}=|\widetilde{M}_{ki}|e^{j\angle{\widetilde{M}_{ki}}},\label{eq:mag_phase}
  \end{align}
\noindent and then sets
  \begin{align}
  \theta=\angle{\widetilde{M}_{ki}},\label{eq:theta}
  \end{align}
\noindent which reduces~\eqref{eq:complex_transformed} to
\begin{subequations}\label{eq:real_transformed}
  \begin{align}
  &\widetilde{M}_{ik}^\prime=|\widetilde{M}_{ik}|,\\
  &\widetilde{M}_{ki}^\prime=|\widetilde{M}_{ki}|,
  \end{align}
 \end{subequations}
\noindent where the fact that $\angle\widetilde{M}_{ki}=-\angle\widetilde{M}_{ik}$ has been used. Note that a complex coupling element can be always transformed into a real coupling element using the transformation matrix in Fig.~\ref{fig:transformation_complex}. In most cases, after successive transformations, the final network  is composed of only real inverters and gyrators, where the minimum number of gyrators is equal to the nonreciprocity order of the transfer function, i.e. the order of $P_{21}(s)/P_{12}(s)$, as shall be shown in the illustrative example of the next section.

\section{Illustrative Example}
\label{sec_example}

\begin{figure}[!t]
  \center
  \psfrag{0}[c][c][0.75]{$0$}
  \psfrag{1}[c][c]{\footnotesize $1$}
  \psfrag{2}[c][c]{\footnotesize $2$}
  \psfrag{S}[c][c]{\footnotesize $S$}
  \psfrag{L}[c][c]{\footnotesize $L$}
  \psfrag{a}[c][c][0.75]{$-0.877+j0.729$}
  \psfrag{b}[c][c][0.75]{$-0.877-j0.729$}
  \psfrag{d}[c][c][0.75]{$-0.877+j0.729$}
  \psfrag{c}[c][c][0.75]{$-0.877-j0.729$}
  \psfrag{g}[c][c][0.75]{$1.140$}
  \psfrag{i}[c][c][0.75]{$1.140$}
  \psfrag{h}[c][c][0.75]{$1.140$}
  \psfrag{j}[c][c][0.75]{$1.140$}
  \psfrag{e}[c][c][0.75]{$-1.806$}
  \psfrag{f}[c][c][0.75]{$1.806$}
  \includegraphics[width=8.6cm]{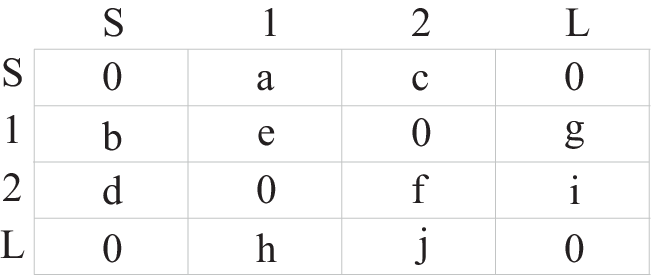}\\
  \caption{Computed coupling matrix corresponding to the transversal array network in Fig.~\ref{fig:transversal}.}\label{fig:cmx_example_transversal}
\end{figure}

Consider the second-order two-port network with transmission and reflection functions
  \begin{subequations}\label{eq:example_trans_ref}
  \begin{align}
  &S_{21}(s)=\dfrac{s-1.5}{0.25s^2+1.3s+1.5052},\\
  &S_{11}(s)=\dfrac{0.25s^2+0.125}{0.25s^2+1.3s+1.5052}.
  \end{align}
 \end{subequations}
According to~\eqref{eq:s_polynomial},
  \begin{subequations}\label{eq:example_polyn_a}
  \begin{align}
  &P_{21}(s)=s-1.5,\\
  &F_{11}(s)=0.25s^2+0.125,\\
  &H(s)=0.25s^2+1.3s+1.5052.
  \end{align}
 \end{subequations}
and, according to~\eqref{eq:mini_condition} with $-$ sign and~\eqref{eq:f11_f22} with $+$ sign due to the even order of $H(s)$ in this case,
 \begin{subequations}\label{eq:example_polyn_b}
  \begin{align}
  &P_{12}(s)=-[(-s)-1.5]=s+1.5,\\
  &F_{22}(s)=0.25(-s)^2-0.125=0.25s^2-0.125,
  \end{align}
 \end{subequations}
so that, from~\eqref{eq:s_polynomial},
  \begin{subequations}\label{eq:example_trans_ref_2}
  \begin{align}
  &S_{12}(s)=\dfrac{s+1.5}{0.25s^2+1.3s+1.5052},\\
  &S_{22}(s)=\dfrac{0.25s^2-0.125}{0.25s^2+1.3s+1.5052}.
  \end{align}
 \end{subequations}
Inserting~\eqref{eq:example_polyn_a} and~\eqref{eq:example_polyn_b} into~\eqref{eq:y_conversion_final} yields then
  \begin{subequations}\label{eq:example_y}
  \begin{align}
  &Y_{21}(s)=\dfrac{-s+1.5}{0.5s^2+1.6302},\\
  &Y_{22}(s)=\dfrac{1.3s}{0.5s^2+1.6302},
  \end{align}
 \end{subequations}
\noindent which, upon factorization in the form~\eqref{eq:y_expansion}, become
  \begin{subequations}\label{eq:example_y_expand}
  \begin{align}
  &Y_{21}(s)=\dfrac{-1-j0.8307}{s-j1.8057}+\dfrac{-1+j0.8307}{s+j1.8057},\\
  &Y_{22}(s)=\dfrac{1.3}{s-j1.8057}+\dfrac{1.3}{s+j1.8057}.
  \end{align}
 \end{subequations}
\noindent The admittance functions~\eqref{eq:example_y_expand} correspond to the transversal array network in Fig.~\ref{fig:transversal} and to the coupling matrix in Fig.~\ref{fig:cmx_transversal} with $N=2$. The entries of the coupling matrix are provided by~\eqref{eq:correspond}, which yield
  \begin{subequations}
  \begin{align}
  &B_1=-1.806,\\
  &B_2=1.806,\\
  &\widetilde{M}_{SL}=0,\\
  &M_{L1}=M_{L2}=\sqrt{1.3}=1.140,\\
  &\widetilde{M}_{S1}=\frac{-1+j0.831}{\sqrt{1.3}}=-0.877+j0.729,\\
  &\widetilde{M}_{S2}=\frac{-1-j0.831}{\sqrt{1.3}}=-0.8771-j0.729,
  \end{align}
 \end{subequations}
\noindent corresponding to the transversal coupling matrix given in Fig.~\ref{fig:cmx_example_transversal}. Note that the coupling between the source and the nodes (resonators) in this matrix are all complex and that the coupling coefficients at transpose locations are conjugate pairs, whereas the coupling between the load and the nodes are all real. This results from the transversal element in Fig.~\ref{fig:element_nonrecip}, where a complex inverter connects the source and the resonator whereas a conventional real inverter connects the load and the resonator.

For practical realizability, one may wish to suppress the coupling between the source and node~2, i.e. eliminate the entries $[1,3]$ and $[3,1]$ in Fig.~\ref{fig:transversal}. This may be done by applying the complex rotation operation given by~\eqref{eq:rotation}, which results into the folded topology shown in Fig.~\ref{fig:example_topo_folded}.

\begin{figure}[!t]
  \center
  \psfrag{0}[c][c][0.75]{$0$}
  \psfrag{1}[c][c]{\footnotesize $1$}
  \psfrag{2}[c][c]{\footnotesize $2$}
  \psfrag{S}[c][c]{\footnotesize $S$}
  \psfrag{L}[c][c]{\footnotesize $L$}
  \psfrag{a}[c][c][0.75]{$-1.240+j1.030$}
  \psfrag{b}[c][c][0.75]{$-1.240-j1.030$}
  \psfrag{e}[c][c][0.75]{$0.331-j1.775$}
  \psfrag{f}[c][c][0.75]{$0.331+j1.775$}
  \psfrag{g}[c][c][0.75]{$0.954+j0.793$}
  \psfrag{h}[c][c][0.75]{$0.954-j0.793$}
  \psfrag{i}[c][c][0.75]{$0.658+j0.793$}
  \psfrag{j}[c][c][0.75]{$0.658-j0.793$}
  \includegraphics[width=8.6cm]{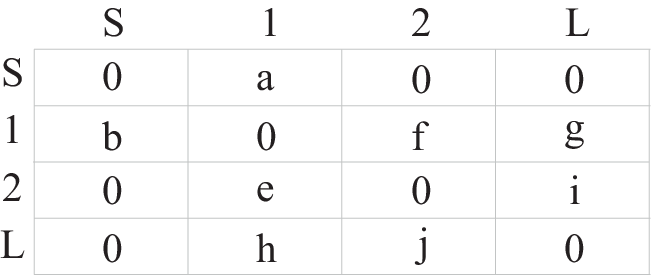}\\
  \caption{Coupling matrix in folded form after topology transformation.}\label{fig:cmx_example_folded}
\end{figure}

\begin{figure}[!t]
  \centering
  \psfrag{1}[c][c]{\footnotesize $1$}
  \psfrag{2}[c][c]{\footnotesize $2$}
  \psfrag{3}[l][c]{\footnotesize Source/Load}
  \psfrag{4}[l][c]{\footnotesize Resonating Node}
  \psfrag{5}[l][c]{\footnotesize Complex Coupling}
  \psfrag{S}[c][c]{\footnotesize $S$}
  \psfrag{L}[c][c]{\footnotesize $L$}
  \includegraphics[width=4cm]{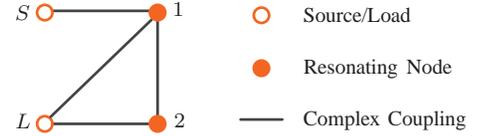}\\
  \caption{Folded topology corresponding to the coupling matrix in Fig.~\ref{fig:cmx_example_folded}.}\label{fig:example_topo_folded}
\end{figure}

Note that all the coupling coefficients are complex after the topology transformation. To transform them into real quantities, one applies the rotation matrix defined in Fig.~\ref{fig:transformation_complex}. First, one may transform the coupling between the source and the first node, i.e. the entry $[1,2]$ in the coupling matrix of Fig.~\ref{fig:cmx_example_folded}. One sets $i=2$ and $\theta=\angle(-1.240+j1.030)=2.448$ in the rotation matrix of Fig.~\ref{fig:transformation_complex} by applying~\eqref{eq:theta}, which results in the coupling matrix and topology shown in Figs.~\ref{fig:cmx_example_folded_2} and~\ref{fig:example_topo_folded_2}, respectively. The coupling between the source and the first node has indeed become a real value, and so has the coupling between the load and the first node. Two complex couplings are still left in Fig.~\ref{fig:example_topo_folded_2}. One may next transform the coupling between the load and the second node, i.e. the entry $[4,3]$ in the coupling matrix of Fig.~\ref{fig:cmx_example_folded_2}. One assigns $i=3$ and $\theta=\angle(0.658-j0.793)=-0.878$ in the rotation matrix of Fig.~\ref{fig:transformation_complex} by applying~\eqref{eq:theta}. The resulting coupling matrix and topology are shown in Figs.~\ref{fig:cmx_example_folded_3} and~\ref{fig:example_topo_folded_3}, respectively. All the couplings except the one between the first node and the second node are now real, and can therefore be implemented using conventional inverters since they represent reciprocal couplings. Also note that the coupling between the first and second nodes is purely imaginary, and hence nonreciprocal which represents a gyrator. Interestingly, only one nonreciprocal coupling is left, which exactly corresponds to the nonreciprocal order\footnote{The nonreciprocal order is defined here as the order of the rational function $S_{21}(s)/S_{12}(s)$. Taking for instance \eqref{eq:example_trans_ref} and \eqref{eq:example_trans_ref_2}, $S_{21}(s)/S_{12}(s)=-(s-1.5)/(s+1.5)$. So the nonreciprocal order is one in this case.} in the scattering functions~\eqref{eq:example_trans_ref} and~\eqref{eq:example_trans_ref_2}.

The magnitude and group delay responses associated with the final coupling matrix given in Fig.~\ref{fig:cmx_example_folded_3}, computed using the classical method presented in~\cite{cameron2007microwave}, are shown in Fig.~\ref{fig:example_mag} and Fig.~\ref{fig:example_gp}, respectively. The calculated responses using the coupling matrix are confirmed to agree with those calculated using the polynomial-based functions in~\eqref{eq:example_trans_ref} and \eqref{eq:example_trans_ref_2}. Moreover, as expected, the magnitude is reciprocal whereas the group delay is nonreciprocal.

\begin{figure}[!t]
  \center
  \psfrag{0}[c][c][0.75]{$0$}
  \psfrag{1}[c][c]{\footnotesize $1$}
  \psfrag{2}[c][c]{\footnotesize $2$}
  \psfrag{S}[c][c]{\footnotesize $S$}
  \psfrag{L}[c][c]{\footnotesize $L$}
  \psfrag{a}[c][c][0.75]{$1.613$}
  \psfrag{b}[c][c][0.75]{$1.613$}
  \psfrag{e}[c][c][0.75]{$-1.389+j1.154$}
  \psfrag{f}[c][c][0.75]{$-1.389-j1.154$}
  \psfrag{g}[c][c][0.75]{$-1.240$}
  \psfrag{h}[c][c][0.75]{$-1.240$}
  \psfrag{i}[c][c][0.75]{$0.658+j0.793$}
  \psfrag{j}[c][c][0.75]{$0.658-j0.793$}
  \includegraphics[width=8.6cm]{cmx_example_folded.eps}\\
  \caption{Coupling matrix after a first-round complex coupling transformation.}\label{fig:cmx_example_folded_2}
\end{figure}

\begin{figure}[!t]
  \centering
  \psfrag{1}[c][c]{\footnotesize $1$}
  \psfrag{2}[c][c]{\footnotesize $2$}
  \psfrag{3}[l][c]{\footnotesize Source/Load}
  \psfrag{4}[l][c]{\footnotesize Resonating Node}
  \psfrag{5}[l][c]{\footnotesize Complex Coupling}
  \psfrag{6}[l][c]{\footnotesize Real Coupling}
  \psfrag{S}[c][c]{\footnotesize $S$}
  \psfrag{L}[c][c]{\footnotesize $L$}
  \includegraphics[width=4cm]{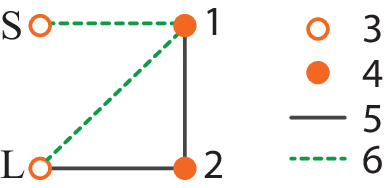}\\
  \caption{Topology corresponding to the coupling matrix in Fig.~\ref{fig:cmx_example_folded_2}.}\label{fig:example_topo_folded_2}
\end{figure}

\begin{figure}[!t]
  \center
  \psfrag{0}[c][c][0.75]{$0$}
  \psfrag{1}[c][c]{\footnotesize $1$}
  \psfrag{2}[c][c]{\footnotesize $2$}
  \psfrag{S}[c][c]{\footnotesize $S$}
  \psfrag{L}[c][c]{\footnotesize $L$}
  \psfrag{a}[c][c][0.75]{$1.613$}
  \psfrag{b}[c][c][0.75]{$1.613$}
  \psfrag{e}[c][c][0.75]{$j1.806$}
  \psfrag{f}[c][c][0.75]{$-j1.806$}
  \psfrag{g}[c][c][0.75]{$-1.240$}
  \psfrag{h}[c][c][0.75]{$-1.240$}
  \psfrag{i}[c][c][0.75]{$1.030$}
  \psfrag{j}[c][c][0.75]{$1.030$}
  \includegraphics[width=8.6cm]{cmx_example_folded.eps}\\
  \caption{Coupling matrix after the second-round complex coupling transformation.}\label{fig:cmx_example_folded_3}
\end{figure}

\begin{figure}[!t]
  \centering
  \psfrag{1}[c][c]{\footnotesize $1$}
  \psfrag{2}[c][c]{\footnotesize $2$}
  \psfrag{3}[l][c]{\footnotesize Source/Load}
  \psfrag{4}[l][c]{\footnotesize Resonating Node}
  \psfrag{5}[l][c]{\footnotesize Imaginary Coupling}
  \psfrag{6}[l][c]{\footnotesize Real Coupling}
  \psfrag{S}[c][c]{\footnotesize $S$}
  \psfrag{L}[c][c]{\footnotesize $L$}
  \includegraphics[width=4cm]{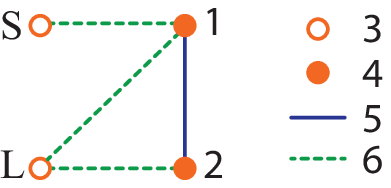}\\
  \caption{Topology corresponding to the coupling matrix in Fig.~\ref{fig:cmx_example_folded_3}.}\label{fig:example_topo_folded_3}
\end{figure}

\begin{figure}[!t]
  \centering
  \psfrag{a}[c][c]{\footnotesize Lowpass Frequency $\Omega$}
  \psfrag{b}[c][c]{\footnotesize Magnitude (dB)}
  \psfrag{c}[l][c]{\footnotesize Coupling}
  \psfrag{e}[l][c]{\footnotesize Matrix}
  \psfrag{d}[l][c]{\footnotesize Polynomial}
  \psfrag{g}[c][c]{\footnotesize $|S_{21}|$ \& $|S_{12}|$}
  \psfrag{h}[r][c]{\footnotesize $|S_{11}|$ \& $|S_{22}|$}
  \includegraphics[width=8.6cm]{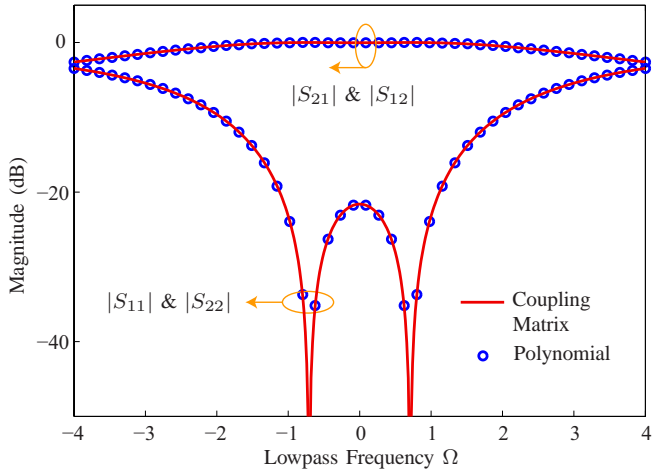}\\
  \caption{Magnitude of the scattering parameters calculated by the coupling matrix in Fig.~\ref{fig:cmx_example_folded_3} and by the polynomial-based functions in \eqref{eq:example_trans_ref} and \eqref{eq:example_trans_ref_2}.}\label{fig:example_mag}
\end{figure}

\begin{figure}[!t]
  \centering
  \psfrag{a}[c][c]{\footnotesize Lowpass Frequency $\Omega$}
  \psfrag{b}[c][c]{\footnotesize Group Delay (s)}
  \psfrag{c}[l][c]{\footnotesize Coupling}
  \psfrag{e}[l][c]{\footnotesize Matrix}
  \psfrag{d}[l][c]{\footnotesize Polynomial}
  \psfrag{h}[r][c]{\footnotesize $\tau_{21}$}
  \psfrag{g}[c][c]{\footnotesize $\tau_{12}$}
  \includegraphics[width=8.6cm]{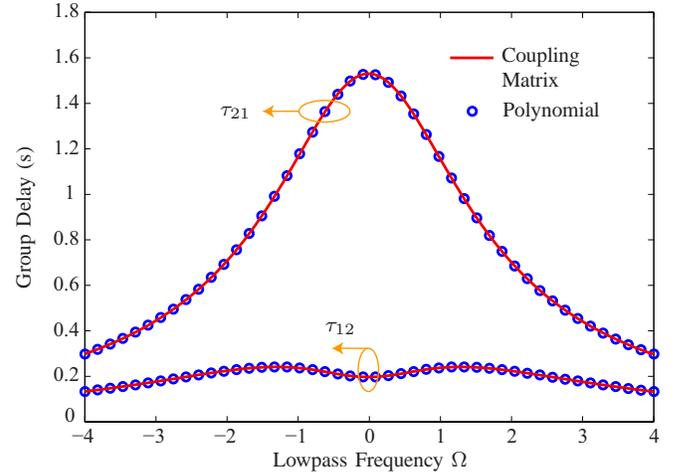}\\
  \caption{Group delay the transfer functions calculated by the coupling matrix in Fig.~\ref{fig:cmx_example_folded_3} and by the polynomial-based functions in \eqref{eq:example_trans_ref} and \eqref{eq:example_trans_ref_2}.}\label{fig:example_gp}
\end{figure}

\section{Further Discussion}
\subsection{Possible Implementation of Gyrators}

There are mainly three ways to practically realize gyrators: using ferromagnetic materials~\cite{gyrator_ferromagnetic}, transistor-based circuits~\cite{gyrator_active_1,gyrator_active_2,gyrator_TL,gyrator_multiport}, and magnetless nonreciprocal metamaterial (MNM) structures~\cite{gyrator_MNM}.

Ferromagnetic gyrators are based on the Faraday rotation effect~\cite{gyrator_ferromagnetic}. This is a mature technology that has been largely commercialized. Such gyrators do not consume power when biased by a permanent magnet, but they are bulky and therefore not suitable for circuit integration. Transistor-based gyrators are fully integrable and therefore are widely used in integrated circuits. They usually exhibit wide bandwidths due to their small size compared with the wavelength. However, they consume significant power and suffer from moderate power handling. MNM-based gyrators would represent a compromise between the other two approaches. They are fully integrable and have medium power handling. Since they may be very thin (metasurfaces), and therefore could easily fit in tiny areas regions between resonators when composed of a single unit cell, they may represent an ideal implementation for nonreciprocal phaser of the type of Fig.~\ref{fig:example_topo_folded_3}.

\subsection{Comparison between Real Inverters and Gyrators}

The gyrator shares several fundamental properties with a real inverter. It has indeed been shown in Sec.~\ref{sec:complex_inverter} that, as a particular case of a complex inverter, it inverts the load impedance, just as a real inverter. The only difference is that the phases of the gyrator for transmissions in the two directions differ by $\pi$, which is why gyrators were also called directional inverters~\cite{gyrator_ferromagnetic}.

Table~\ref{tab:comparison} compares the real inverter and the gyrator. Note that the transfer admittance is purely reactive for the real inverter whereas it is purely resistive for the gyrator. Also note that the inversion factors are the same despite the different reciprocity operation.

Note that the series connection of two gyrators produces a reciprocal transformer~\cite{tellegen1948gyrator} similarly to the connection of two real inverters. This property may be used for conversion between gyrators and real inverters.

\begin{table}[!t]
\caption{Comparison between a real inverter and a gyrator ($M$ is a real quantity).}
\makegapedcells
\setcellgapes{3pt}
\begin{tabular}{p{0.8cm}<{\centering}|p{1.8cm}<{\centering}|p{1.8cm}<{\centering}|c|p{1.1cm}<{\centering}}
  \hline
  \hline
    & Y Matrix & ABCD Matrix & \shortstack{Inversion\\Factor} & Reciprocity \\
  \hline
  \shortstack{Real\\Inverter} & $\left[ {\begin{array}{*{20}{c}}
  0&{jM} \\
  {jM}&0
\end{array}} \right]$ & $\left[ {\begin{array}{*{20}{c}}
  0&{\frac{j}{M}} \\
  {jM}&0
\end{array}} \right]$ & $M^2$ & Yes \\
\hline
  Gyrator & $\left[ {\begin{array}{*{20}{c}}
  0&{M} \\
  {-M}&0
\end{array}} \right]$ & $\left[ {\begin{array}{*{20}{c}}
  0&{\frac{1}{M}} \\
  {M}&0
\end{array}} \right]$ & $M^2$ & No \\
  \hline
  \hline
\end{tabular}\label{tab:comparison}
\end{table}

\subsection{Alternative Implementation of Complex Inverters}

Although complex inverters can be transformed into real inverters using the technique described in Sec.~\ref{sec:comp_trans}, they may also be directly implemented by combining real inverters and gyrators, as shall be shown here.

The admittance matrix of a complex inverter [Eq.~\eqref{eq:inverter_admittance}] can be decomposed into two parts,
  \begin{equation}\label{eq:complex_decomposition}
  j\left[ {\begin{array}{*{20}{c}}
  {{0}}&{{\widetilde{M}}}\\
  {{\widetilde{M}^*}}&{{0}}
  \end{array}} \right]=\left[ {\begin{array}{*{20}{c}}
  {{0}}&{{jM_r}}\\
  {{jM_r}}&{{0}}
  \end{array}} \right]+\left[ {\begin{array}{*{20}{c}}
  {{0}}&{{M_i}}\\
  {{-M_i}}&{{0}}
  \end{array}} \right],
  \end{equation}
  \noindent where
  \begin{subequations}\label{eq:real_imag_complex}
  \begin{align}
  &M_r=\text{Re}\left\{\widetilde{M}\right\},\\
  &M_i=-\text{Im}\left\{\widetilde{M}\right\}.
  \end{align}
 \end{subequations}

\begin{figure}[!t]
  \centering
  \psfrag{1}[c][c]{\footnotesize \shortstack{Complex\\Inverter}}
  \psfrag{3}[c][c]{\footnotesize \shortstack{Real\\Inverter}}
  \psfrag{2}[c][c]{\footnotesize Gyrator}
  \includegraphics[width=8.6cm]{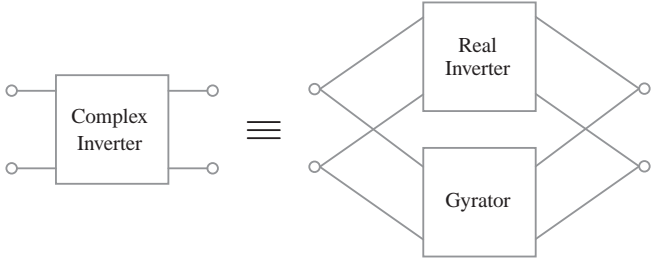}\\
  \caption{Implementation of a complex inverter by connnecting a real inverter and a gyrator in parallel.}\label{fig:complex_implementation}
\end{figure}

Equation~\eqref{eq:complex_decomposition} reveals that a complex inverter can be obtained by parallel-connecting a real inverter and a gyrator, as shown in Fig.~\ref{fig:complex_implementation}. Although this implementation is more complicated than that discussed in Sec.~\ref{sec:comp_trans}, it may represent an alternative solution.

\section{Conclusions}

The coupling matrix synthesis technique has been generalized to the case of nonreciprocal lossless two-port networks. Complex inverter have been introduced to build the nonreciprocal transversal network corresponding to the nonreciprocal coupling matrix. The resulting network has then been transformed into a folded topology through complex similarity transformation. A new similarity transformation technique has also been provided to transform the complex inverters into real inverters and gyrators for implementation simplicity. A second-order example has then been given to illustrate the proposed theory. Finally, a discussion has been provided about possible gyrator implementations, comparison between real inverters and gyrators, and an alternative implementation of complex gyrators.

Beyond its theoretical interest, the extension of the coupling matrix synthesis technique to non-reciprocal networks may find applications in radio analog signal processing systems and various microwave systems involving nonreciprocal components.

%

\bibliographystyle{IEEEtran}
\bibliography{mybib}

\end{document}